\documentclass[useAMS,usenatbib]{mn2e}
\usepackage[paperwidth=215.9mm,paperheight=279.4mm,totalwidth=169mm,totalheight=243mm]{geometry}
\usepackage[dvips]{graphicx} \usepackage{pslatex}
\usepackage{amssymb,amsmath} \usepackage{color,psfrag}
\usepackage{verbatim,xspace} \usepackage{import,afterpage}
\usepackage{supertabular} \usepackage[varg]{txfonts}

\newcommand{\ro}[1]{\ensuremath{\textrm{#1}}}
\newcommand{\kmsec}{\ensuremath{~\ro{km}~ \ro{s}^{-1}}\xspace}

\newcommand{\cm}{\ensuremath{~\ro{cm}^{-2}}}
\newcommand{\Msol}{\ensuremath{M_{\odot}}\xspace}
\newcommand{\lya}{\ensuremath{\ro{Ly}\alpha}\xspace}
 
\newcommand{\hi}{\ensuremath{\ro{H\textsc{i}}}\xspace}
\newcommand{\nhi}{\ensuremath{N_{\ro{H\textsc{i}}}}\xspace}
\newcommand{\dhi}{\ensuremath{n_{\ro{H\textsc{i}}}}\xspace}

\newcommand{\df}{\ensuremath{~ \ro{d} }}
\newcommand{\dd}{\ensuremath{\ro{d} }}

\newcommand{\cN}{\ensuremath{\mathcal{N}}\xspace}

\defcitealias{2008ApJ...681..856R}{R08}
\defcitealias{2009MNRAS.397..511B}{BH09}
\defcitealias{2010MNRAS.403..870B}{BH10}

\title[The bias of DLAs]{The bias of DLAs at $z\sim 2.3$: contraining stellar feedback in shallow potential wells}

\author[Barnes \& Haehnelt]{Luke A. Barnes$^{1}$\thanks{E-Mail:
 L.Barnes@physics.usyd.edu.au (LAB)} and Martin
 G. Haehnelt$^{2}$\footnotemark[1] \\ $^1$SuperScience Fellow, Sydney
 Institute for Astronomy, University of Sydney, 44-70 Rosehill
 Street, Redfern, Australia\\ $^2$Institute of Astronomy and Kavli
 Institute for Cosmology, Madingley Road, Cambridge, CB3 0HA \\ }

\begin{document}

\date{not yet submitted}

\pagerange{\pageref{firstpage}--\pageref{lastpage}} \pubyear{2010}

\maketitle 

\label{firstpage}

\begin{abstract} 
We discuss the recent Baryon Oscillation Spectroscopic Survey 
measurement of a rather high bias factor
for the host galaxies/haloes of Damped Lyman-alpha Absorbers (DLAs), in
the context of our previous modelling of the physical properties of
DLAs within the $\Lambda$ cold dark matter paradigm. Joint modelling of the column
density distribution, the velocity width distribution of associated low
ionization metal absorption, and the bias parameter suggests that DLAs
are hosted by galaxies with dark matter halo masses in the range $10 < \log
M_v < 12$, with a rather sharp cutoff at the lower mass end, 
corresponding to virial velocities of $\sim 35 \kmsec$. The observed
properties of DLAs appear to suggest efficient (stellar) feedback
in haloes with masses/virial velocities below the cutoff and a large
retained baryon fraction ($\ga 35 \%$) in haloes above the cutoff.
\end{abstract}



\section{Introduction} 

Lyman alpha (\lya), seen in absorption in the spectra of quasars, is the most
sensitive method for detecting baryons at high redshift
\citep[e.g.][]{1998ARA&A..36..267R}. \lya absorbers are
classified according to their neutral hydrogen column density,
\nhi. \lya forest absorbers have $\nhi < 10^{17} \cm$, making them
optically thin to ionising radiation. Lyman limit systems (LLS) have
$10^{17} \cm < \nhi < 10^{20.3} \cm$. Damped Lyman alpha absorbers
(DLAs) are the highest column density systems, with $\nhi > 10^{20.3}
\cm$ and have long been known to probe sightlines passing through the
interstellar medium (ISM) of high-redshift galaxies
\citep{2005ARA&A..43..861W}. Direct observations of
the stellar emission of DLA host galaxies are made 
difficult by the overwhelmingly bright background QSO, meaning
that their precise nature has remained controversial
\citep{1997ApJ...487...73P,2000ApJ...536...36K,
2007A&A...468..587C,2012MNRAS.424L...1K}. 
Some consensus has
been reached that the absorption cross-section-selected DLA host
galaxies are generally less massive than typical spectroscopically
confirmed emission-selected galaxies at the same redshift
\citep{1999MNRAS.305..849F,
2000ApJ...534..594H,2001ApJ...559L...1S,2008ApJ...683..321F,
2008MNRAS.390.1349P,2013arXiv1308.2598B,2013arXiv1310.3317R}. 


The most important observational properties of DLAs can be summarised
by their distribution of column density (\nhi), velocity
width ($v_w$) from associated low ionization metal absorbers,
metallicity, and redshift. These properties have proven challenging for
models of DLAs to reproduce, especially the velocity width distribution
of low ionization metal absorption. Many simulations which otherwise
account very well for both DLA properties and for galaxy properties
today struggle to produce enough large-velocity width DLAs
\citep{2008ApJ...683..149R,2008MNRAS.390.1349P,
2009MNRAS.397..411T,2010arXiv1008.4242H}.

\begin{figure} \centering
	\includegraphics[width=0.45\textwidth]{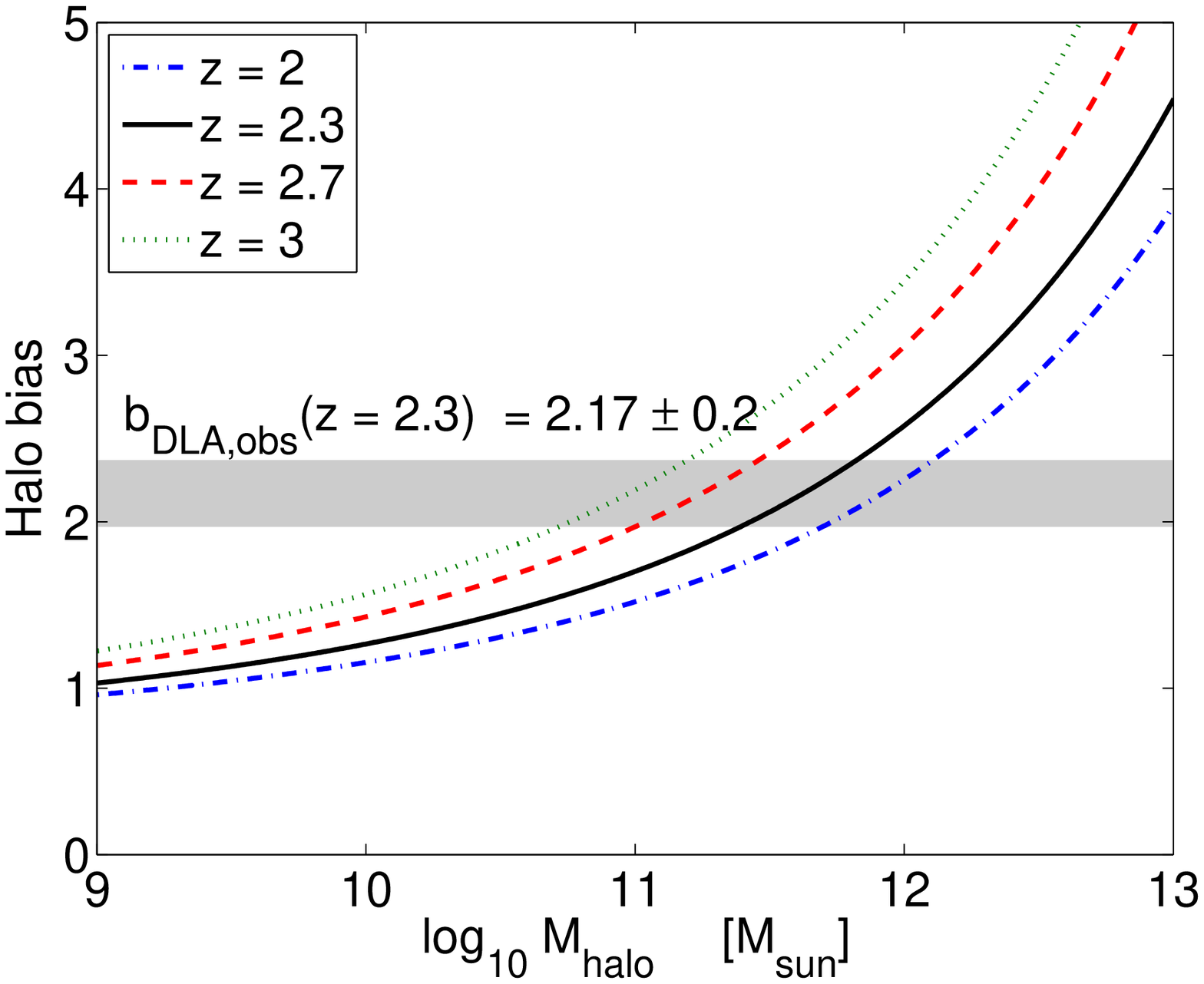}
	\caption{The bias of dark matter haloes, calculated following
 \citet{2002MNRAS.329...61S} for a range of redshifts 
 including the mean redshift of the BOSS bias data,
 $\langle z \rangle = 2.3$. The observed bias and its 1
 $\sigma$ error are shown as a grey shaded region, suggesting
 a typical DLA mass scale of $\sim 10^{11.5}
 \Msol$.} \label{fig:biasM}
\end{figure}

The most comprehensive DLA survey to date comes from the Baryon
Oscillation Spectroscopic Survey \citep[BOSS][]{2013AJ....145...10D},
which is part of the Sloan Digital Sky Survey III
\citep[SDSS III][]{2011AJ....142...72E}. The full sample, based on SDSS Data
Release 9, contains over 150,000 quasar spectra over the redshift range
$2.15 < z < 3.5$ and has discovered 6,839 DLAs, which is an order of
magnitude larger than SDSS II. In particular, BOSS has for the first
time estimated the bias of DLA host galaxies $(b_{\rm DLA})$ with
respect to the matter distribution by cross-correlating DLA absorption
with \lya forest absorption \citep{2012JCAP...11..059F}. The surprisingly
large value of $b_\ro{DLA} = (2.17 \pm 0.20) ~ \beta_F^{0.22}$, where 
$\beta_F \approx 1$ is the \lya forest
distortion parameter, provides an important 
constraint on the distribution of the host
halo masses of the DLA population. Figure \ref{fig:biasM} shows the
bias of dark matter haloes as a function of halo mass, for a range of
redshifts including the mean redshift of the BOSS bias data, $\langle
z \rangle = 2.3$. The measured value of the DLA bias suggests a
typical DLA halo mass of $\sim 10^{11.5} \Msol$, significantly larger
than is found in many simulations \citep[e.g.][]{2008MNRAS.390.1349P,
2013arXiv1310.3317R}.

In \citet{2009MNRAS.397..511B,2010MNRAS.403..870B} we proposed a
simple model for DLAs that simultaneously accounts for their
absorption properties, and also reproduces the emission properties of
a population of very faint \lya emitters observed by
\citet{2008ApJ...681..856R}. Here we revisit our model to see whether
it can also account for the observed DLA bias. In Section \ref{s:DLAmodel} we describe our model for the DLA population. Section \ref{s:obs}
compares our modelling to observations. Section \ref{s:metal}
uses our model to place constraints on the mass-metallicity relation
of DLAs, and compares this relation to the corresponding relation
for luminosity-selected galaxies. In Section \ref{s:Discussion} we discuss 
our results and give our conclusions.

\section{The DLA Model} \label{s:DLAmodel}
In this section, we will summarise our model presented in
\citet{2010MNRAS.403..870B}, and discuss how we calculate the DLA
bias. 
The cosmological parameters assumed here are $(h,
~\Omega_M, ~\Omega_b, ~\Omega_{\Lambda}, ~\sigma_8, ~n, ~Y_p) = (0.71,
~0.281, ~0.0462, ~0.719, ~0.8, ~0.963, ~0.24)$. Note that the
cosmological parameters have been updated from those used in
\citet{2010MNRAS.403..870B} and that in particular the value of
$\sigma_8$ is significantly smaller (0.8 versus 0.9). 

The number density of dark matter haloes is calculated using the
Press-Schechter formalism, with the elliptical-collapse ansatz of
\citet{2002MNRAS.329...61S}. The number of dark matter haloes per unit
comoving volume at redshift $z$ with mass (baryonic + CDM) in the
interval $(M_v,M_v+\dd M_v )$ is estimated as,
\begin{equation}\label{eq:nst}
n_{M_v}(M_v,z) \df M_v = A \left( 1 + \frac{1}{\nu'^{2q}} \right)
\sqrt{\frac{2}{\pi}} \frac{\rho_{0}}{M_v} \frac{\dd\nu'}{\dd M_v} \exp
\left(-\frac{\nu'^2}{2} \right) \df M_v ~,
\end{equation}
where $\sigma_M$ is the rms fluctuation amplitude of the cosmic
density field in spheres containing mass $M_v$, $\rho_{0}$ is the
present cosmic matter density (baryonic + CDM), $\nu' = \sqrt{a}\nu$,
$\nu = \delta_c / [D(z) \sigma_M]$. $D(z)$ is the growth factor at
redshift $z$ \citep{1992ARA&A..30..499C}, $\delta_c = 1.686$, $a =
0.707$, $A \approx 0.322$ and $q = 0.3$. We have used the fitting
formula in \citet{1999ApJ...511....5E} to calculate the matter power
spectrum.

We assign baryons to a given dark matter halo according to its total
mass. The mass of \hi $(M_\hi)$ in a galactic halo is assumed to
scale with the total virial mass $(M_v)$ for large haloes, while being
suppressed for smaller haloes due to the combined effect of
photoionisation from the UV background, galactic winds and perhaps other
feedback processes. This suppression is necessary to avoid
overpredicting the number of DLAs with small velocity width
\citep{1998ApJ...495..647H,2000ApJ...534..594H}. In our model,
\begin{equation} \label{eq:MHI}
M_\hi = f_\hi ~ f_\ro{H,c} ~ \exp \left[ -
 \left(\frac{v_\ro{v,0}}{v_{\ro{v}}} \right)^{\alpha_{\rm e}} \right]
M_v
\end{equation}
where $f_\ro{H,c} = (1 - Y_p)\Omega_b / \Omega_m$ is the cosmic
hydrogen mass fraction; $f_\hi$ is the mass fraction of \hi in haloes,
relative to cosmic; $v_{\ro{v}}$ is the halo virial
velocity\footnote{The virial velocity, virial mass and virial radius
 are related as \citep[e.g.][]{2004MNRAS.355..694M},:
\begin{align}
v_{\ro{v}} &= 96.6 \kmsec ~ \left( \frac{\Delta_{\ro{v}} \Omega_M
 h^2}{24.4} \right)^{\frac{1}{6}} \left( \frac{1+z}{3.3}
\right)^{\frac{1}{2}} \left( \frac{M_v}{10^{11} \Msol}
\right)^{\frac{1}{3}} \label{eq:vc} \\
R_{\ro{v}} & = 46.1 \ro{kpc} ~ \left(
\frac{\Delta_{\ro{v}} \Omega_M h^2}{24.4} \right)^{-\frac{1}{3}}
\left( \frac{1+z}{3.3} \right)^{-1} \left( \frac{M_v}{10^{11} \Msol}
\right)^{\frac{1}{3}} \label{eq:Rc}
\end{align}
where $\Delta_{\ro{v}}$ is the mean overdensity of the halo
\citep[see][]{1998ApJ...495...80B}.}, $v_\ro{v,0}$ is the virial
velocity below which the \hi fraction is suppressed, and $\alpha_{\rm
 e}$ is a parameter which determines the sharpness of the
suppression.

To calculate the DLA cross-section, we need to model the distribution
of neutral gas in the halo. Following the simulations of
\citet[][Equation (9)]{2004MNRAS.355..694M}, we alter the NFW profile
\citep{1996ApJ...462..563N} to give the halo gas a core at $\simeq
3 r_\ro{s}/4$, where $r_\ro{s}$ is the scale radius of the NFW
profile,
\begin{equation}
\rho_\hi(r) = \frac{\rho_0 r_s^3}{(r + \frac{3}{4}r_\ro{s})(r +
 r_\ro{s})^2} ~,
\end{equation}
where $\rho_0$ normalises the profile so that the mass inside the
virial radius is equal to $M_\hi$ as specified by Equation
\eqref{eq:MHI}; see Equation (9)-(11) of
\citet{2004MNRAS.355..694M}. This spherically-symmetric distribution
can be thought of as an effective average profile for a given halo
mass.

The \hi density as a function of radius is specified by the total mass
of the halo $M_v$ and the concentration parameter $c_{\rm v}
\equiv r_{\rm v}/r_{\rm s}$ of the \hi. For the dependence of the
concentration parameter on the mass, we take the mean value of the
$c_{\rm v}-M_v$ correlation for dark matter as given by
\citet{2007MNRAS.378...55M},
\begin{equation}
c_{\rm v} = c_0 \left( \frac{M_v}{10^{11} \Msol}
\right)^{-0.109} \left( \frac{1+z}{4} \right)^{-1} ~.
\end{equation} 
For dark matter, \citet{2007MNRAS.378...55M} found that $c_0 \approx
3.5$, with a log-normal distribution and a scatter around this mean
value of $\Delta (\ln c_{\rm v}) = 0.33$, in agreement of the results
of \citet{2001MNRAS.321..559B} and \citet{2002ApJ...568...52W}. As in
\citet{2010MNRAS.403..870B}, we will find later that a significantly
larger $c_0$ is required for the baryons; we will use the column
density distribution of DLAs to constrain $c_0$. The gas in the DLAs 
can be expected to self-shield against the
meta-galactic ionizing UV background. The corresponding
self-shielding radius in the DM haloes we are studying here is
generally smaller than the virial radius. We therefore set the outer
radius of the \hi to be the virial radius. Given the number density
of \hi atoms $\dhi = \rho_\hi / m_H$, we can use the relationship
between impact parameter $b$ and column density on a line of sight
through the system,
\begin{equation}
\nhi (b) = 2 \int_0^{\sqrt{r_v^2 - b^2}} \dhi \left( r = \sqrt{b^2 +
 y^2} \right) \df y ~,
\end{equation}
to calculate the DLA cross-section of a given halo,
\begin{equation}
\sigma_\ro{DLA} = \pi b_\ro{DLA}^2 \quad \ro{where} \quad \nhi
(b_\ro{DLA}) = N_\ro{DLA} \equiv 10^{20.3} \ro{cm}^{-2} ~.
\end{equation}
Note that in reality the incidence for DLA absorption is unlikely to
have unit covering factor within a given radius; thus the DLA
cross-section calculated should be considered as an effective average
DLA cross-section for haloes of given mass/virial velocity. See, for example, 
\citet{2013arXiv1308.2598B} for more detailed modelling 
of the spatial distribution and kinematics of the gas contributing 
to the DLA cross-section.

We can now calculate the column density distribution, defined such
that the number of systems ($\dd^2 \cN$) intersected by a random line
of sight between absorption distance\footnote{The absorption distance
 is defined by \begin{equation} \dd X \equiv \frac{H_0} {H(z)}
 (1+z)^2 \df z ~.
\end{equation}.}
$X$ and $X+\dd X$, with \hi column density between $N_{\hi}$ and
$N_{\hi}+\dd N_{\hi}$ is, 
\begin{align}
\dd^2 \cN &= f(\nhi,X) \df X \df \nhi \\ \Rightarrow \quad f(\nhi,X)
&= \frac{c}{H_0} \int n_{M_v}(M_v,X) \left| \frac{\dd \sigma}{\dd \nhi}
(N_{\hi}|M_v,X) \right| \df M_v .
\end{align}

The velocity width $v_w$ of a DLA is defined by
\citet{1997ApJ...487...73P} in their pioneering survey as the velocity
interval encompassing 90\% of the total integrated optical depth. Given
the conditional probability distribution of $v_w$ given $v_{\ro{v}}$, $p(v_\ro{w} |
v_{\ro{v}}) \df v_\ro{w}$, we can calculate the distribution of DLA
velocity width along a random line of sight per unit absorption
distance,
\begin{equation} \label{eq:lvw}
l(v_\ro{w},X) = \frac{c}{H_0} \int	p(v_\ro{w} | v_{\ro{v}}(M_v)) ~
n_{M_v}(M_v,X) ~ \sigma_{\ro{DLA}}(M_v,X) \df M_v, ~.
\end{equation}
The distribution $p(v_\ro{w} | v_{\ro{c}})$ is chosen to have
lognormal form in $x_v \equiv v_w/v_{\ro{v}}$,
\begin{equation} \label{eq:pvwid}
p(x_v) = \frac{1}{x_v \sqrt{2\pi} \sigma} \exp \left( {-\frac{(\ln x_v
 - \mu)^2}{2\sigma^2}} \right),
\end{equation}
where we parameterize the distribution using the peak $x_\ro{peak}$
and full width half maximum\footnote{For a lognormal
 distribution, \begin{align} \label{eq:lognorm} \sigma &= \frac{1}{\sqrt{2\ln 2}}\sinh^{-1}
 \frac{x_\ro{FWHM}} {2 x_\ro{peak}} \\ \mu &= \ln x_\ro{peak} +
 \sigma^2 \end{align}} $x_\ro{FWHM}$. In previous works, we have
used $p(v_\ro{w} | v_{\ro{c}})$, drawn from the simulations of
\citet{2008MNRAS.390.1349P}. This distribution is approximately fit by
a lognormal distribution with $x_\ro{peak} \approx 0.61$, $x_\ro{FWHM}
\approx 0.45$. A lognormal distribution is also found by
\citet{2013ApJ...769...54N} to fit the observed velocity width
distribution. As we will see later, we need to leave $x_\ro{peak}$ and
$x_\ro{FWHM}$ as free parameters in order to simultaneously
reproduce the velocity width distribution and the DLA bias parameter
$b_{\rm D}$.

Finally, we calculate the DLA bias $b_{\rm D}(z)$ in our model as,
\begin{equation} \label{eq:bDLA}
b_{\rm D}(z) = \frac{\int_0^\infty b_{\rm h}(M,z) ~ n_{M_v}(M_v,z) ~
 \sigma_\ro{\rm DLA}(M_v,z) ~ \df M_v}{\int_0^\infty n_{M_v}(M_v,z) ~
 \sigma_\ro{\rm DLA}(M_v,z) ~ \df M_v} ~,
\end{equation}
where $b_{\rm h}(M_v,z)$ is the bias of dark matter haloes as a function
of mass and redshift. The halo bias is calculated using the
ellipsoidal collapse model of \citet{2002MNRAS.329...61S}, and shown
in Figure \ref{fig:biasM}.

\begin{figure} \centering
	\includegraphics[width=0.45\textwidth]{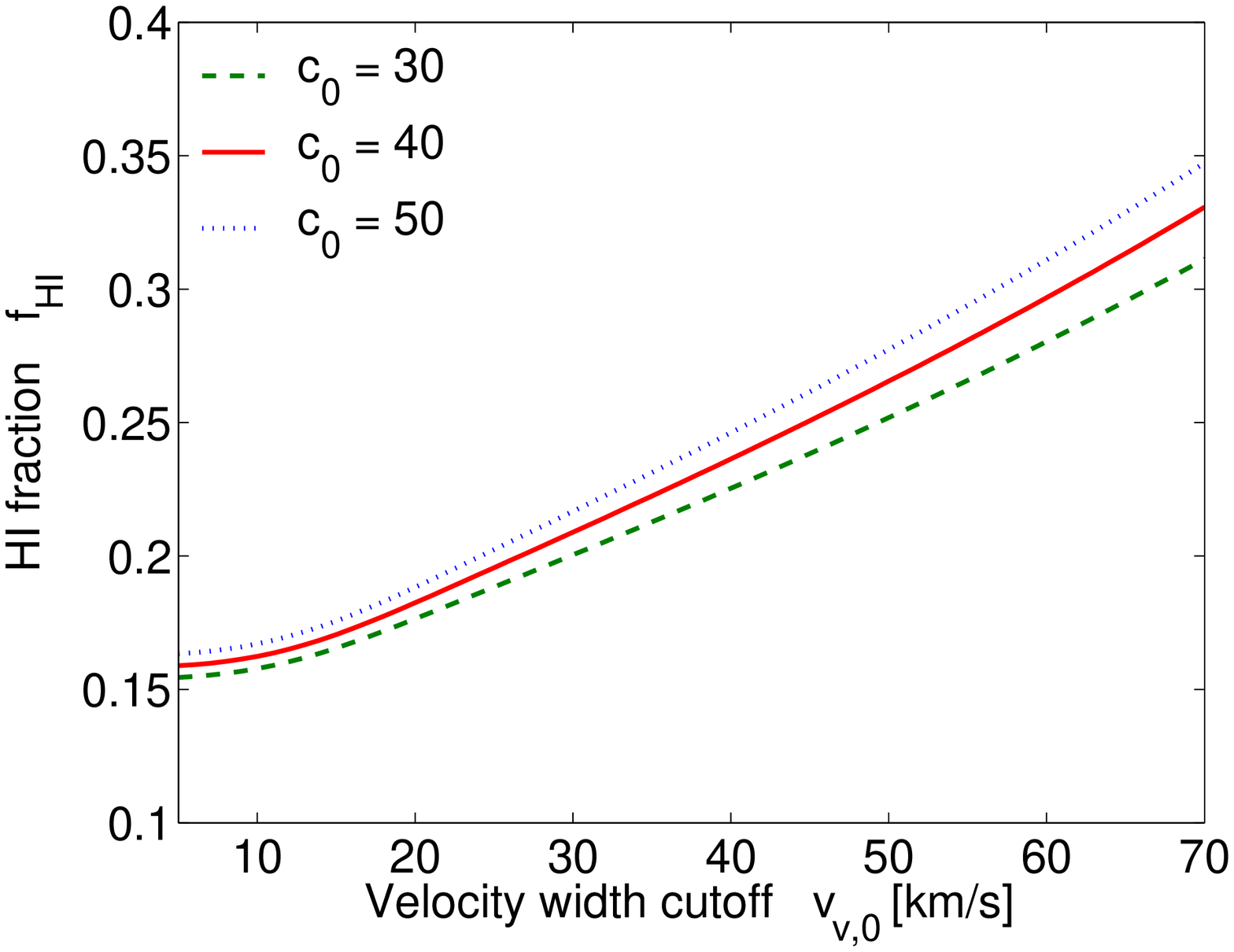}
	\caption{The fraction of HI in a halo as compared to cosmic
 $f_{\hi}$, from Equation \eqref{eq:MHI}, is adjusted to hold
 $l_{\ro{DLA}} \equiv \dd \cN / \dd X$ constant. This plot
 shows the relationship between $f_{\hi}$ and the
 normalisation of the \hi concentration-mass relation
 $c_0$, and the low-mass suppression of \hi in haloes
 $v_\ro{v,0}$.} \label{fig:fe_vs}
\end{figure}

Finally, the mean redshift of observations is slightly different for the
column density, velocity width and bias samples. This is taken into
account in the model. The free parameters of the model are then
the fraction of \hi in the halo relative
to cosmic $f_{\hi}$, the normalisation of the \hi concentration-mass
relation $c_0$, the location $v_\ro{v,0}$ and sharpness $\alpha_{\rm
 e}$ of the low-mass suppression of \hi, and the two parameters
constraining the peak and width of the distribution of DLA velocity
width. We will constrain these parameters using the \hi column density
distribution, the DLA velocity width distribution, and the DLA bias. In
particular, we will hold $\dd \cN / \dd X =
\int_{N_{\ro{DLA}}} f(\nhi,X) \df \nhi$ constant for all the models,
which effectively fixes $f_{\hi}$. The relationship between $f_{\hi}$
and the other parameters of the model is shown in Figure
\ref{fig:fe_vs}. The fiducial values for each of these parameters,
which will be justified in the next section, are as follows,
\begin{align}
v_\ro{v,0} &= 35 \kmsec ~,&&\alpha_{\rm e} = 3, \label{eq:fidmodel}\\
\dd \cN / \dd X &=0.08 ~, &&c_0 = 40, \\
x_\ro{peak} &= 0.7 ~, &&x_\ro{FWHM} = 0.8. 
\end{align}

\section{Comparison to Observations} \label{s:obs}
We will first show how the model accounts for the column density and
velocity width distribution of DLAs, and the constraints that are
placed on the parameters of the model.

\begin{figure*} \centering
	\begin{minipage}{0.45\textwidth}
		\includegraphics[width=\textwidth]{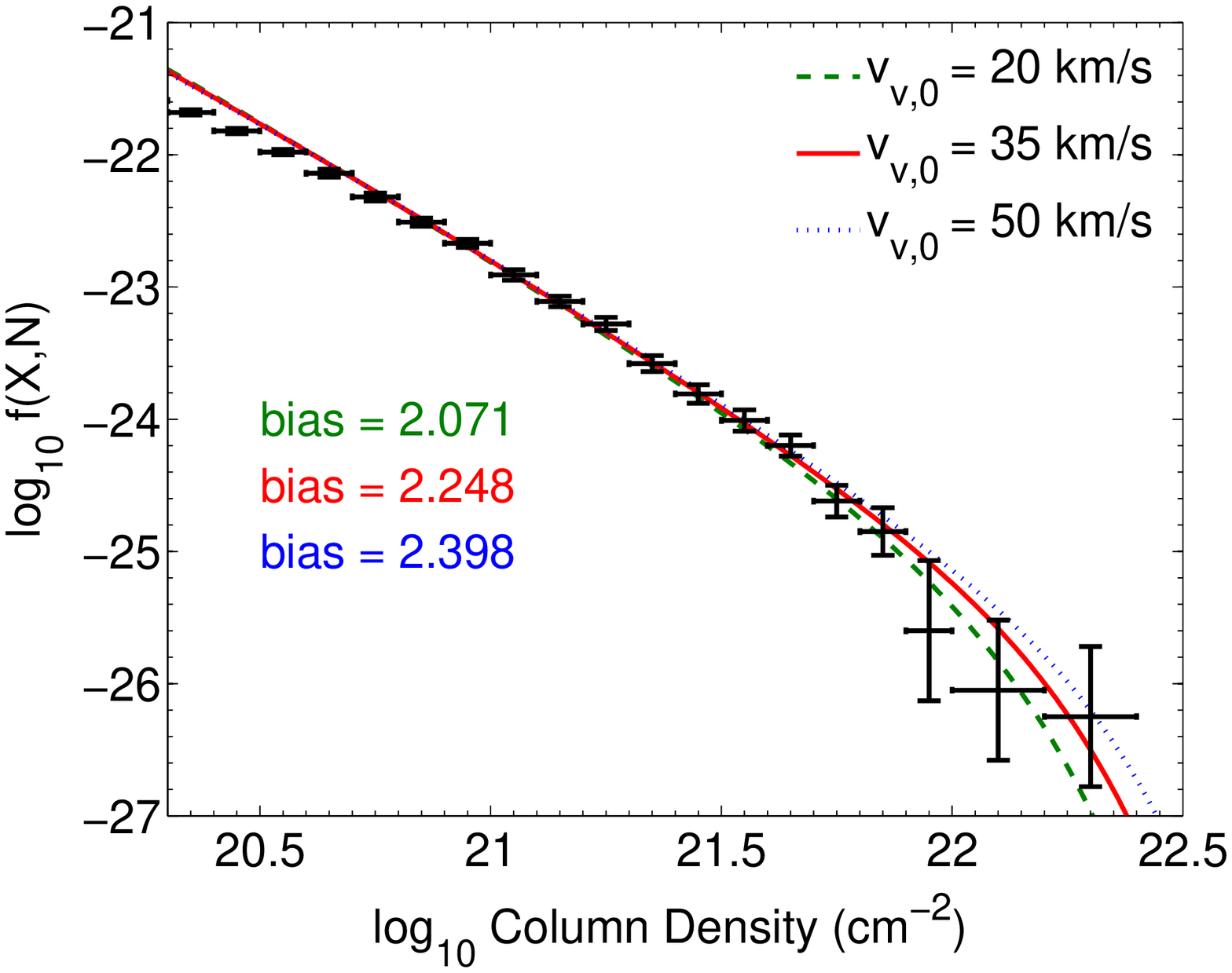}
	\end{minipage}
	\begin{minipage}{0.45\textwidth}
		\includegraphics[width=\textwidth]{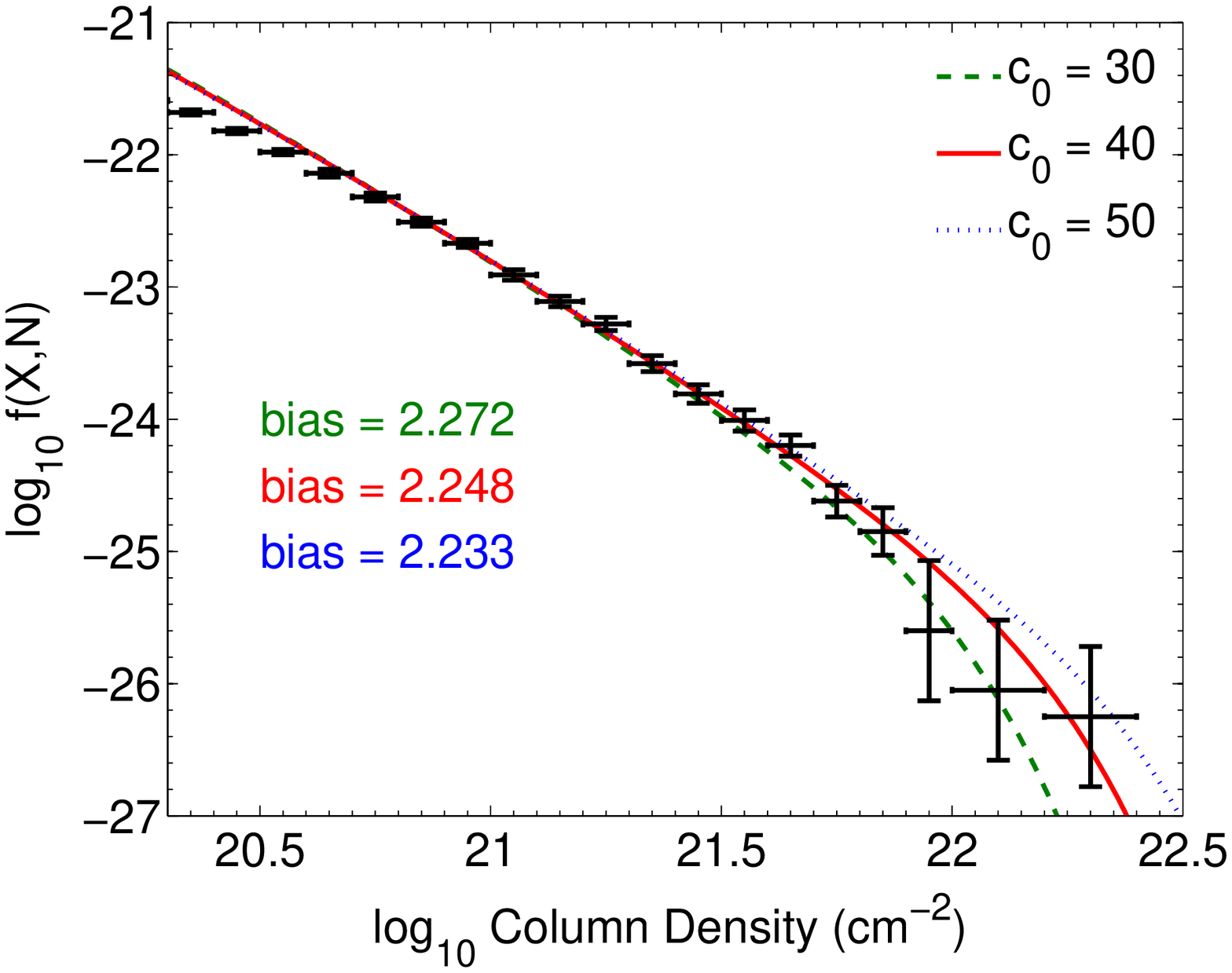}
	\end{minipage}
	\begin{minipage}{0.45\textwidth}
		\includegraphics[width=\textwidth]{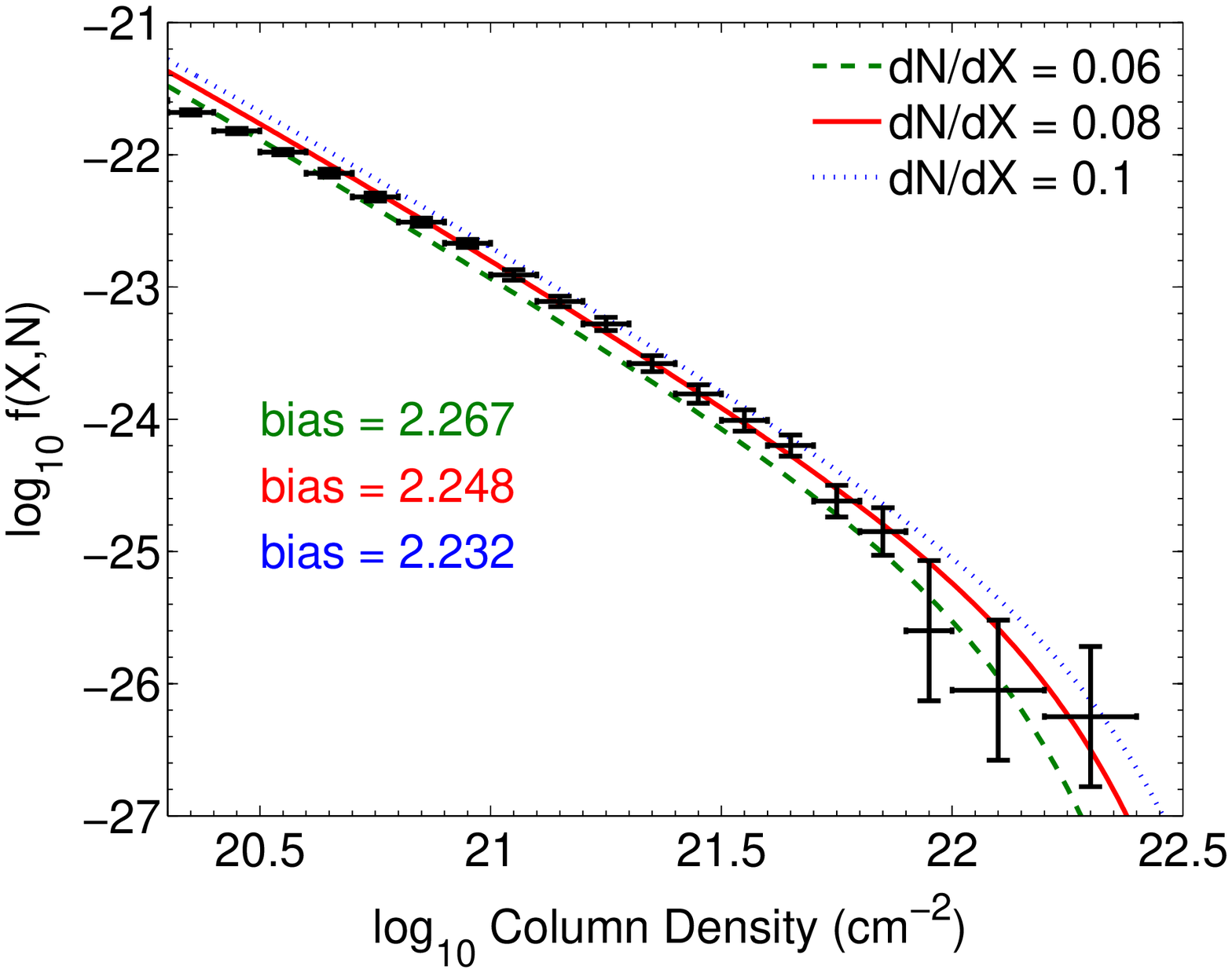}
	\end{minipage}
	\hspace{0.05\textwidth}
	\begin{minipage}{0.40\textwidth}
			\caption{The \hi column density distribution
 $f(X,N)$ for changes in the parameters of
 the model, together with the data of
 \citet{2012AandA...547L...1N}. The fiducial
 model has parameters $v_{v,0} = 35 \kmsec$,
 $c_0 = 40$, and $\dd \cN / \dd X = 0.08$,
 shown by the solid red line in each panel. The
 legend shows which parameter of the fiducial
 model has been changed. The bias is also
 shown, colour coded. The redshift assumed in
 the model is the mean redshift of the data,
 $\langle z \rangle = 2.5$.} \label{fig:fX}
 	\end{minipage}
\end{figure*}

Figure \ref{fig:fX} shows the \hi column density distribution for
changes in the parameters of the model, together with the data of
\citet{2012AandA...547L...1N}. The redshift assumed in the model is
the mean redshift of the data, $\langle z \rangle = 2.5$. The fiducial
model (solid red in each panel) is a good fit to the data. There is a slight
overprediction at low \nhi, most likely due to the flattening of the
observed $f(X,N)$ as we approach the LLS regime in which
photoionisation effects are relevant \citep{2013MNRAS.430.2427R};
 we have neglected such effects
in our modelling. Note that our chosen value of $\dd \cN / \dd X$ is
slightly higher than observed \citep{2009ApJ...696.1543P} to take this
into account.

The effect of $v_{v,0}$ on $f(X,N)$ is largely due to the dependence
of $f_\hi$ on $v_{v,0}$. As $v_{v,0}$ increases, gas is removed from
low mass haloes, and thus $f_\hi$ must increase to hold $\dd \cN / \dd
X$ constant. The top right panel shows that the high \nhi turnover of
$f(X,N)$ constrains $c_0$ to about 25\%. Increasing the concentration
of the \hi in a given halo increases the maximum \nhi for sightlines
passing through the halo. This puts higher \nhi in smaller, more
abundant haloes, boosting the high \nhi end of the distribution
function. The bottom left panel justifies the choice of $\dd \cN / \dd
X = 0.08$, the other lines passing under or over almost all of the
data.

\begin{figure*} \centering
	\begin{minipage}{0.45\textwidth}
		\includegraphics[width=\textwidth]{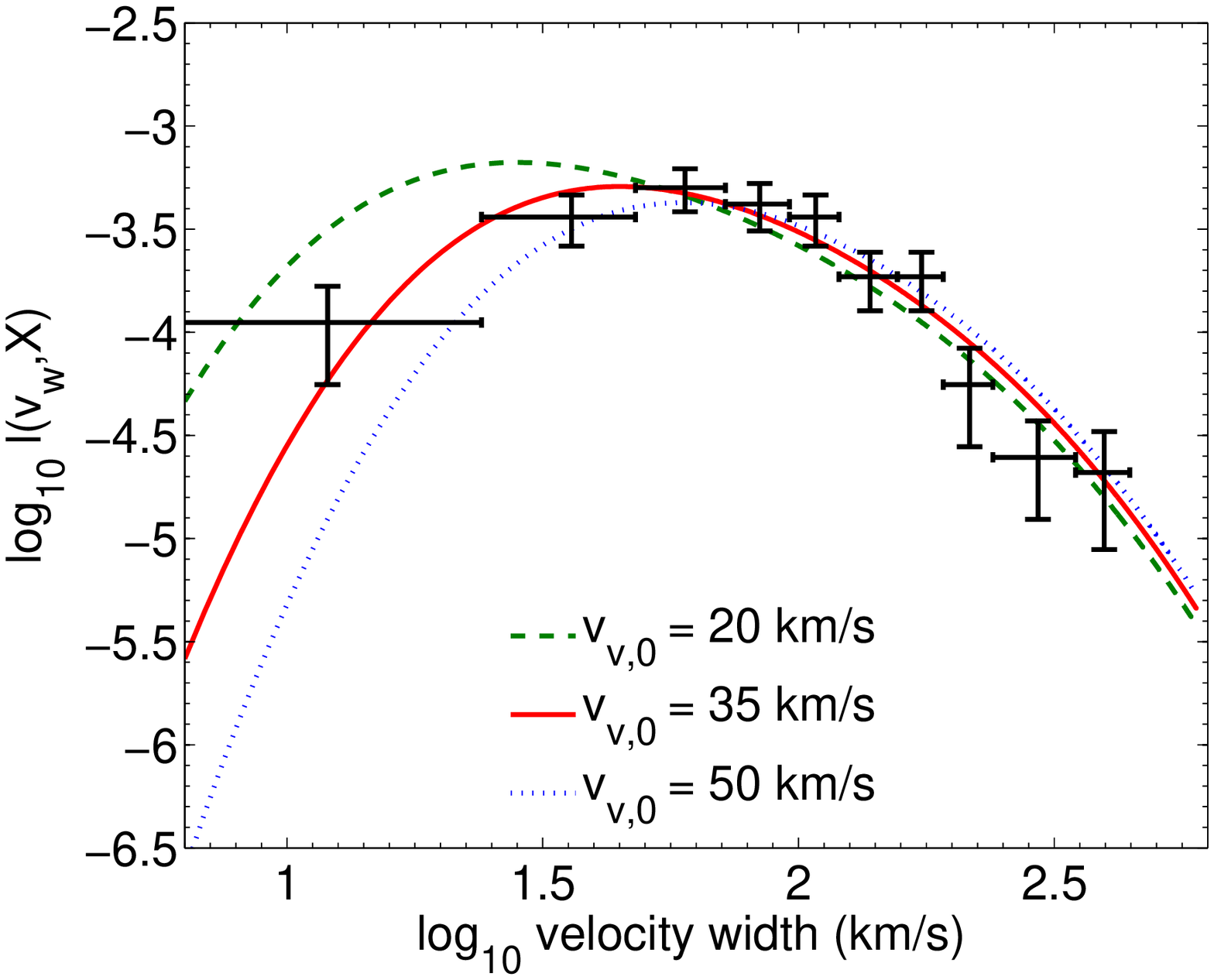}
	\end{minipage}
	\begin{minipage}{0.45\textwidth}
		\includegraphics[width=\textwidth]{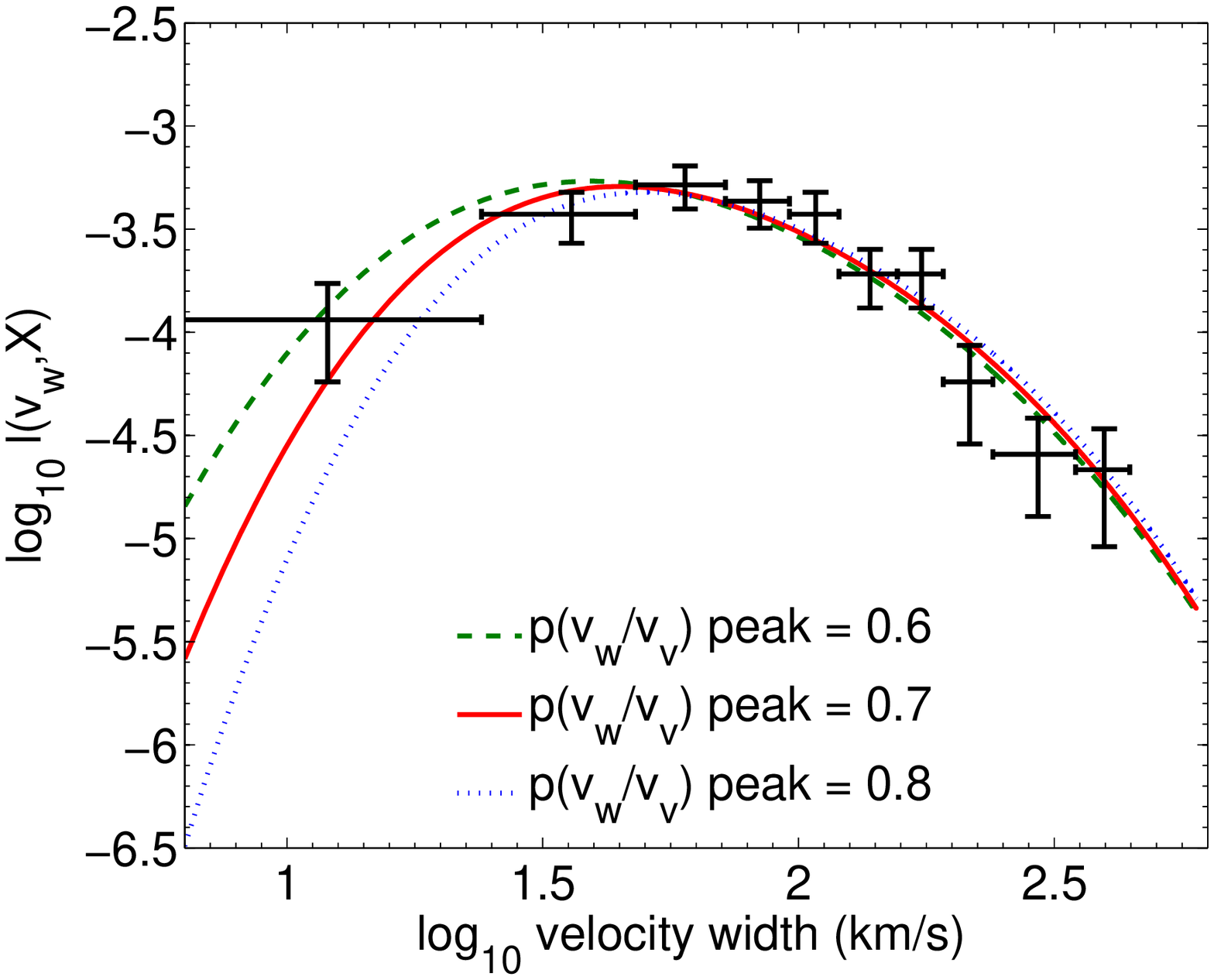}
	\end{minipage}
	\begin{minipage}{0.45\textwidth}
		\includegraphics[width=\textwidth]{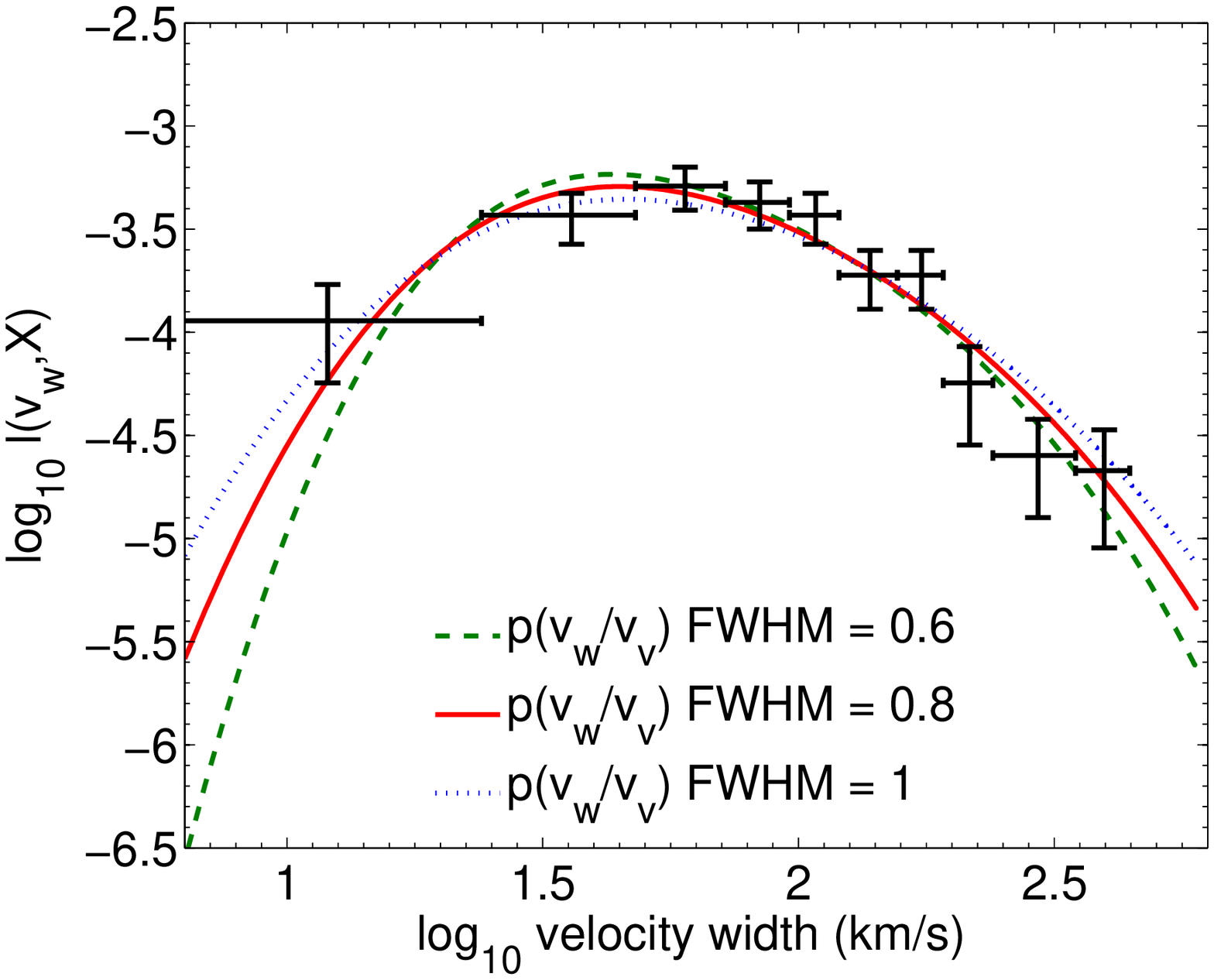}
	\end{minipage}
	\begin{minipage}{0.45\textwidth}
		\includegraphics[width=\textwidth]{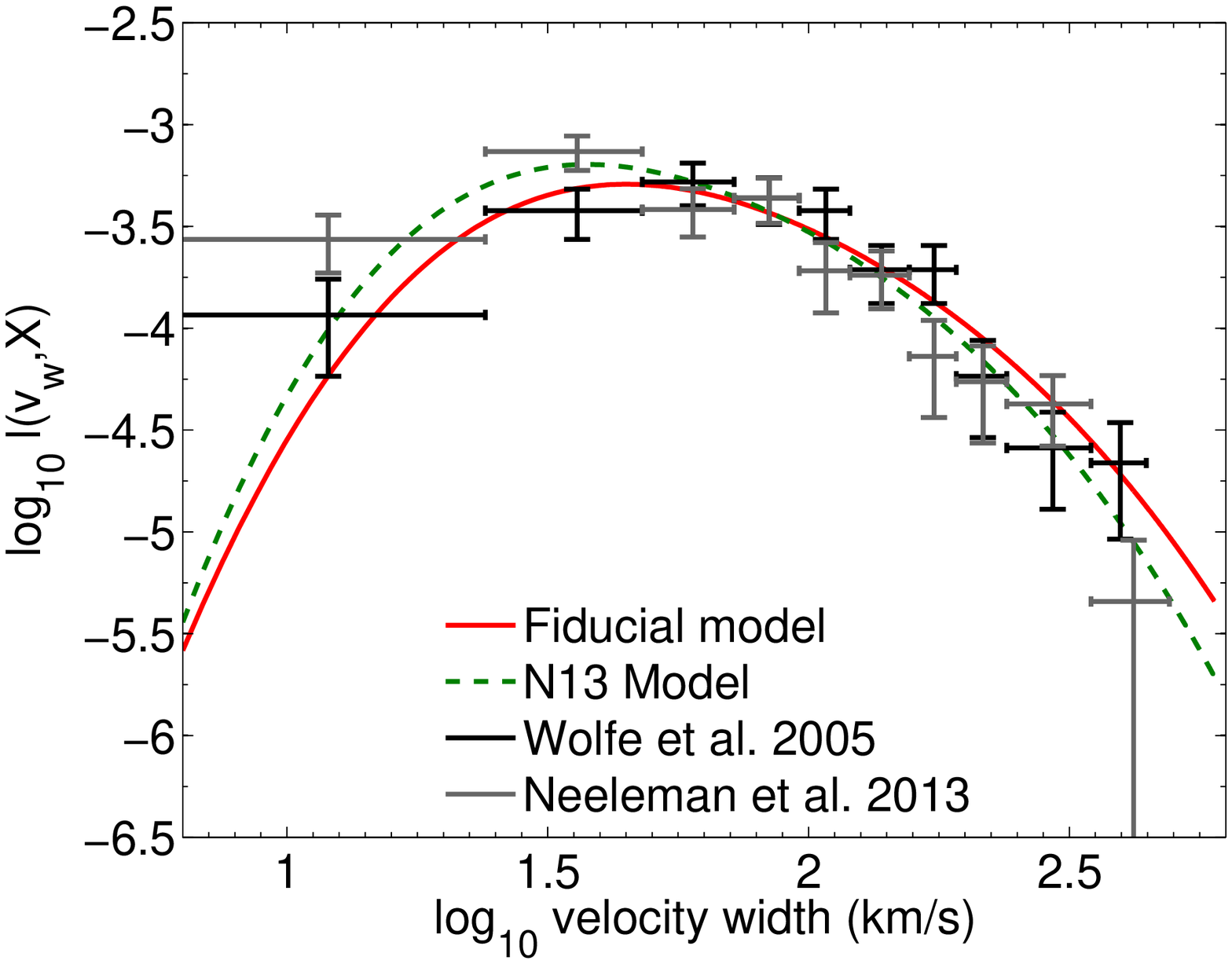}
	\end{minipage}
	\caption{The effect of changing modelling parameters
 on the DLA velocity width distribution $l(v_w,X)$. The black
 crosses show the data from \citet{2005ARA&A..43..861W}. 
 The redshift assumed in the model is the mean redshift of the data,
 $\langle z \rangle = 3$; \citet{2008MNRAS.390.1349P} and
 \citet{2013ApJ...769...54N} note that there is very little
 redshift evolution. The lower right panel compares the data set of 
 \citet{2005ARA&A..43..861W} with \citet{2013ApJ...769...54N}, and
 shows our best fit model for each. The best fit model to the 
 Neeleman et al. data (green dashed line, ``N13 Model'')
 has a conditional velocity width distribution $p(v_\ro{w} | v_{\ro{c}})$
 that is narrower and peaks at a lower value than the fiducial model 
 $x_\ro{peak} = 0.6$, $x_\ro{FWHM} = 0.6$. The difference between 
 the two data sets is further discussed in an appendix.}
 \label{fig:vw}. 
\end{figure*}

Figure \ref{fig:vw} shows the observed velocity width distribution $l(v_w,X)$ of low
ionization absorption of DLAs from \citet{2005ARA&A..43..861W}, along with the
effects of changing our model parameters. Again, our fiducial model is
in red. The redshift assumed in the model is the mean redshift of the
data, $\langle z \rangle = 3$; \citet{2008MNRAS.390.1349P} and
\citet{2013ApJ...769...54N} note that there is very little redshift
evolution. Our fiducial model is again a reasonable fit to the data.

The top left panel shows that $v_{v,0}$ has a dramatic effect on the
velocity width distribution, particularly the low end. This is
expected, as the distribution $p(v_\ro{w} | v_{\ro{c}})$ in Equation
\eqref{eq:lvw} directly ties the velocity width to the virial
velocity. Note that changing $c_0$ has a very small effect on
$l(v_w,X)$, and changing $\dd \cN / \dd X$ scales each line
vertically.

The top right and bottom left panels show the effect of altering the
peak and FWHM of the velocity width distribution. The parameter
$x_\ro{peak}$ has a similar effect to $v_{v,0}$, shifting the
distribution to larger/smaller velocity widths. Its value is constrained to be
$\sim 0.7$. Likewise, the data constrains the width of the distribution to be $\sim 0.8$. 

The bottom right panel also shows in light grey the velocity width
distribution data of \citet{2013ApJ...769...54N}. Note that there is a
considerable difference between this data set and that of
\citet{2005ARA&A..43..861W} at the low velocity width end. 
We discuss this further in an appendix. The solid red line shows our 
best-fit model for the Wolfe et al.
compilation. The dashed green line shows that the Neeleman et al. data, which
has significantly more low-velocity systems, can be instead be fit by a model
with $x_\ro{peak} = 0.6$ and $x_\ro{FWHM} = 0.6$. The difference between
the fiducial and N13 model will not affect our conclusions.

We turn now to the model's prediction of the DLA bias. BOSS has
measured the DLA bias to be $b_\ro{DLA} = (2.17 \pm 0.20) ~
\beta_F^{0.22}$. The bias is shown in Figure \ref{fig:fX}, colour
coded. The redshift assumed in our model is the mean redshift of the
data, $\langle z \rangle = 2.3$. If we use our best fit model from previous
papers, which assumed that $v_{v,0} = 50$\kmsec, the bias is slightly overpredicted. 

\begin{figure} \centering
	\includegraphics[width=0.45\textwidth]{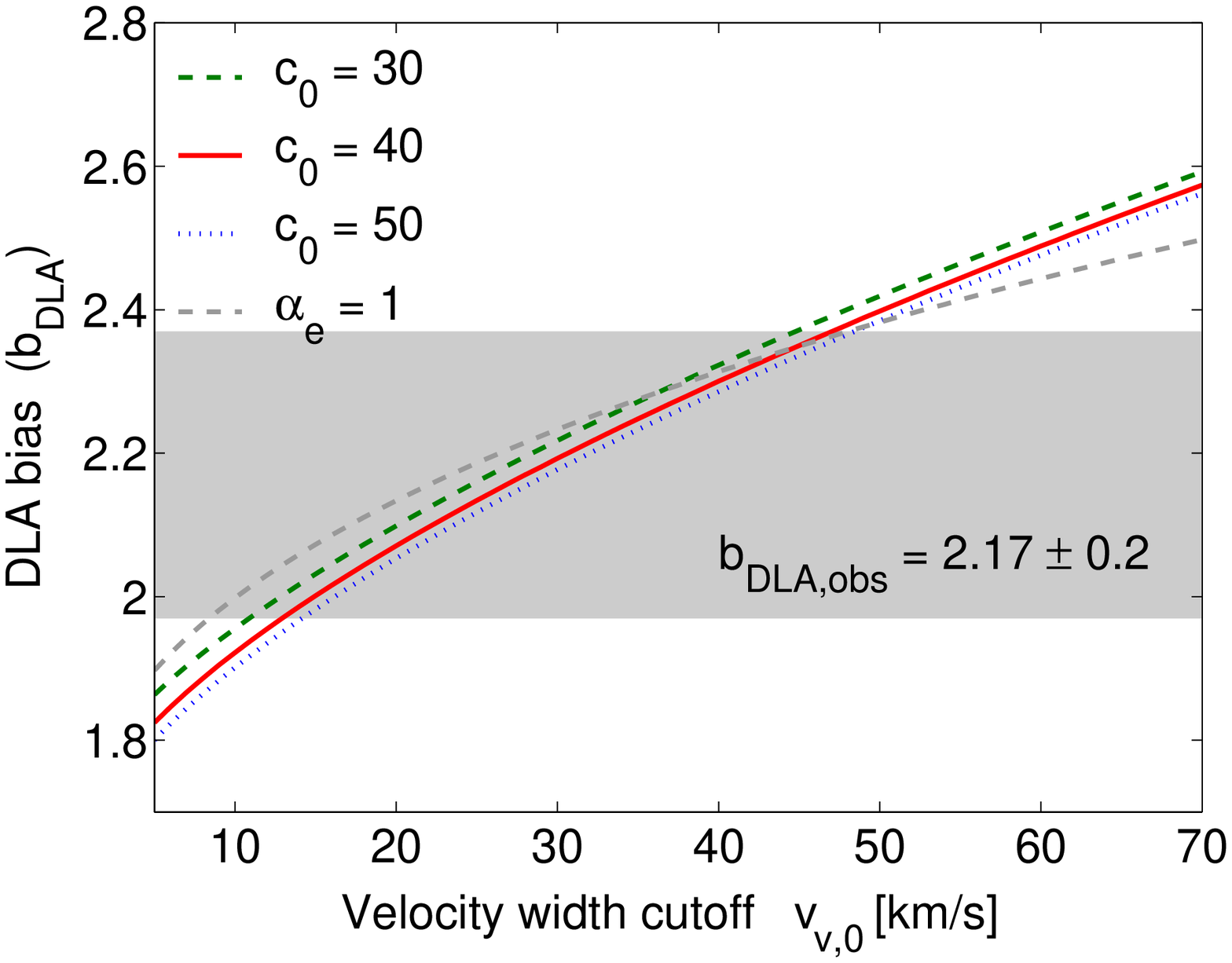}
	\caption{The dependence of the DLA bias $(b_{\rm DLA})$ on the
 parameter $v_{v,0}$, which denotes the virial velocity below
 which the \hi mass in haloes is exponentially
 suppressed. The gray shaded region shows the $\pm 1 \sigma$
 measurement of BOSS: $b_{\rm DLA} = (2.17 \pm 0.20) ~\beta_F^{0.22}$,
 assuming $\beta_F = 1$. The fiducial model is in red, and the other lines 
 show the effect of altering $c_0$ and $\alpha_{\rm e}$. Note that altering
 $\dd \cN / \dd X$ has very little effect on the predicted bias.}
 \label{fig:biasVv0}
\end{figure}

Our preferred value is $v_{v,0} = 35$\kmsec; a range of parameters is shown in Figure
\ref{fig:biasVv0}. The figure shows that varying $c_0$ over 
a range consistent with the
the column density and velocity width distributions does little to
change our prediction of the DLA bias; changing $\dd \cN / \dd X$ 
within the limits shown in Figure \ref{fig:fX} has an even smaller effect. 
The grey dashed line
in Figure \ref{fig:biasVv0} shows the effect of changing the sharpness
of the \hi suppression from $\alpha_{\rm e} = 3$ (so that the \hi
content of galaxies is suppressed $\propto \exp (-\ro{const} /
M_v)$) to $\alpha_{\rm e} = 1$ ($\propto \exp (-\ro{const} /
v_\ro{v})$). The effect on the value of $v_{v,0}$ inferred from
the observed bias is small.

\begin{figure}
 \centering
	\includegraphics[width=0.45\textwidth]{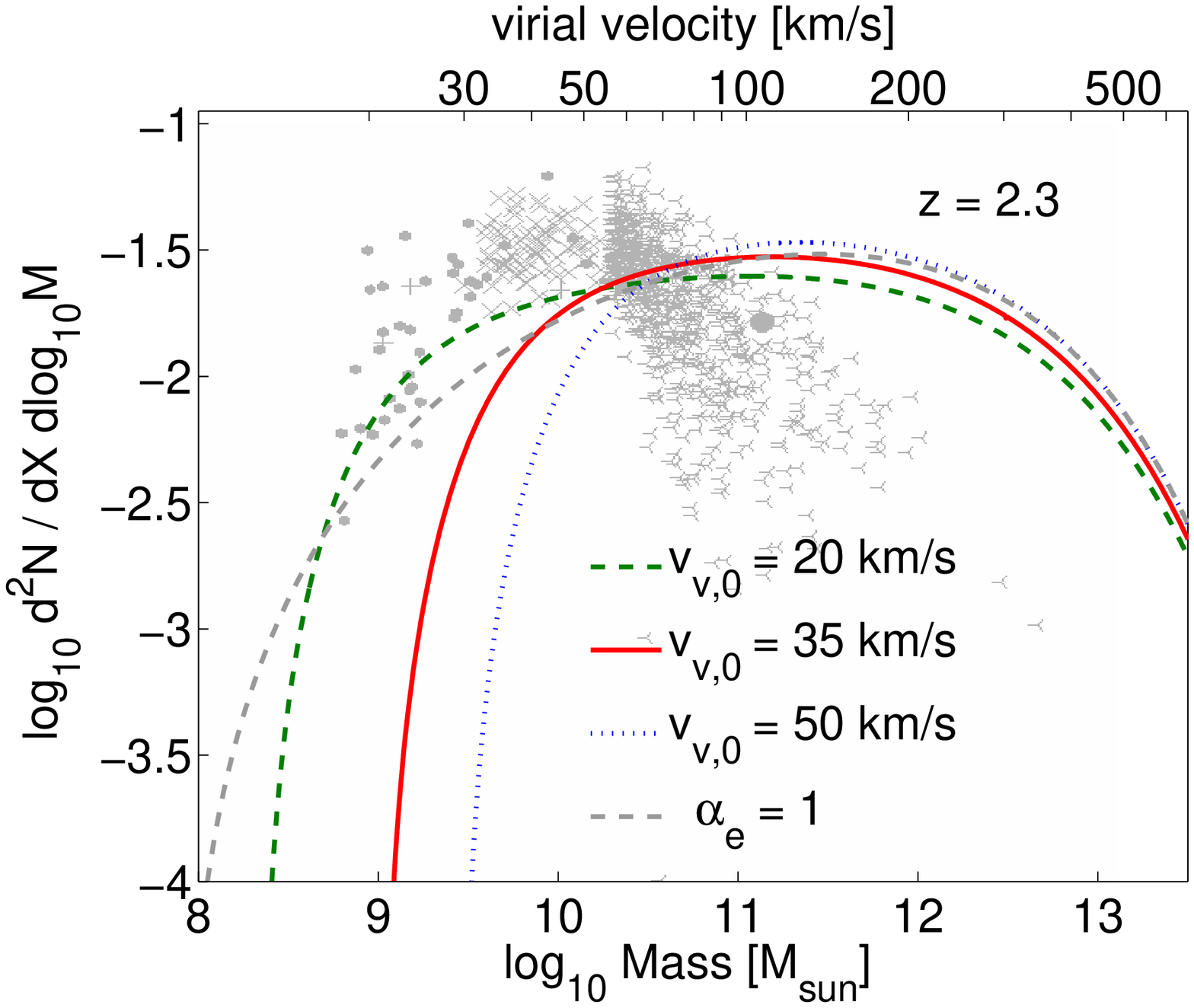}
	\caption{The contribution of different mass ranges (or, equivalently, virial
 velocity ranges) to the incidence rate of DLAs ($\dd^2 N /
 \df X \df \log_{10} M_v$) at $z = 2.3$. The fiducial model is in red, the solid lines show 
 the effect of changing $v_{v,0}$, and the dashed line shows the effect of 
 setting $\alpha_e = 1$. Also shown in grey are the results of the
 cosmological simulation of \citet{2008MNRAS.390.1349P}.}
 \label{fig:dNdXM}
\end{figure}

For our fiducial model, the 
contribution to the DLA incidence rate (Figure \ref{fig:dNdXM}) 
has a broad peak from $10^{10} \Msol$ to $10^{12} \Msol$, a mass range that is 
just below the peak of the 
stellar mass to total halo mass obtained by dark matter halo
abundance matching analyses for the stellar mass function of galaxies 
\citep{2013MNRAS.428.3121M,2013ApJ...762L..31B}.

Mechanical and thermal feedback from supernovae is generally assumed to be the
mechanism by which accretion and star-formation is suppressed in low
mass haloes $M_v < 10^{11} \Msol$. Note that our fiducial
model requires that the peak of the velocity width distribution of DLAs
$x_\ro{peak}$ in a halo of given virial velocity to be slightly larger and the distribution
to be wider than has been found in e.g. the simulations of DLAs
by \citet{2008MNRAS.390.1349P}.

Figure \ref{fig:dNdXM} shows the mass distribution of the DLA incidence rate, 
\begin{equation} \label{eq:dNdXM}
\frac{\dd^2 \cN}{\dd X ~ \dd M_v} = \frac{c}{H_0} ~ n_{M_v}(M_v,X) ~ \sigma_{\ro{DLA}}(M_v,X) ~,
\end{equation}
where the fiducial model is in red, the solid lines show the effect of changing 
$v_{v,0}$, the dashed line shows the effect of setting $\alpha_e = 1$, and the
grey points show the DLAs of the cosmological simulation
 of \citet{2008MNRAS.390.1349P}. The figure shows clearly that at
the low mass end the reduction of the DLA cross-section has to extend
to more massive haloes and therefore deeper potential
wells than in the \citet{2008MNRAS.390.1349P} simulations. Our
combined analysis of velocity width distribution and DLA bias
therefore provides important additional clues/constraints for how
stellar feedback operates in low mass haloes/shallow potential wells.

\section{DLA Metallicity} \label{s:metal}

We can extend our model to constrain the DLA mass-metallicity relationship. 
We assume a probability distribution for metallicity $[M/H]$,
given halo mass $M_v$ and velocity width $v_w$, that is Gaussian with a
dispersion $\sigma_{[M/H]}$ and a mean $[M/H]-v_w,M_v$ relationship that we 
will write as,
\begin{equation} \label{eq:Zmassvw}
[M/H]_\ro{mean} = \alpha_M \log \left( \frac{M_v}{10^{11} \Msol} \right) ~+~ \alpha_v \log \left( \frac{v_w}{\bar{x} ~ v_{\ro{v}}(M_v)} \right)
 ~+~ \beta_M ~,
\end{equation}
where $\bar{x} = \ro{e}^{\mu + \frac{1}{2}\sigma^2}$ is the mean of the lognormal 
distribution (Equation \ref{eq:pvwid}). This form of the relationship 
assumes that halo mass/virial velocity is the main controlling 
factor for average DLA properties, but allows for the possibility
that DLAs with above average velocity width 
(and thus probably above average star formation rate) have above
average metallicity. The mean DLA $[M/H]-M_v$
relation, averaging over $v_w$, has the form,
\begin{equation} \label{eq:Zmass}
[M/H]_\ro{mean} = \alpha_M \log \left( \frac{M_v}{10^{11} \Msol} \right) 
 ~+~ \beta_M ~,
\end{equation}
with the same parameter values as in Equation \eqref{eq:Zmassvw}.

To compare with observations, we calculate the joint probability 
distribution for $[M/H]$ and $v_w$,
\begin{equation} \label{eq:dNdXZ}
p([M/H],v_w) = \int p([M/H] | v_w M_v) p(v_w|v_{\ro{v}}(M_v)) p(M_v) \df M_v ~,
\end{equation}
where the first term in the integral is the normal distribution mentioned above, 
the second term is the lognormal velocity width distribution of Equation \eqref{eq:pvwid},
and the third term is the mass distribution of Equation \eqref{eq:dNdXM}, normalized to 
unity.

\begin{figure*} \centering
	\includegraphics[width=0.8\textwidth]{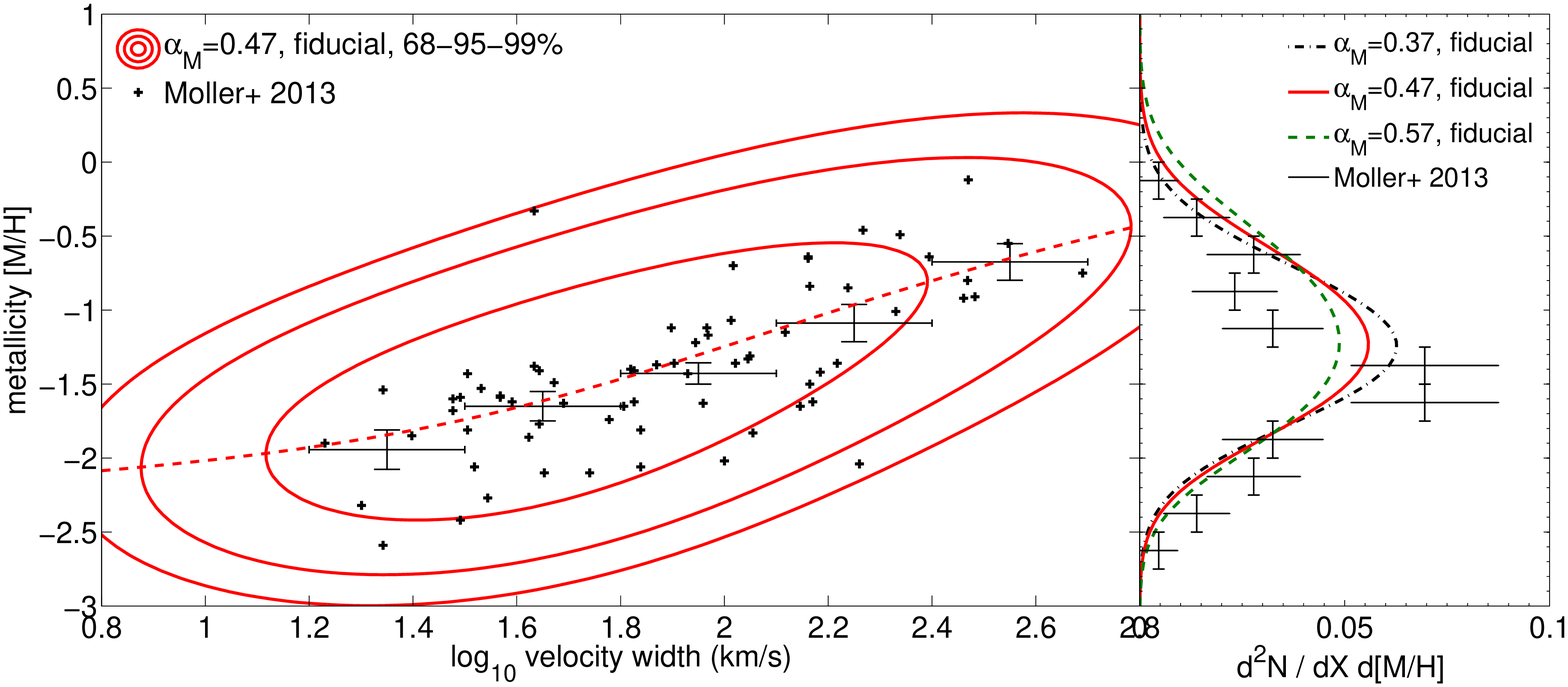}
	\caption{\emph{Left}: 68-95-99\% contours in the $[M/H] - v_w$ plane, 
	along with the data of \citet{2013MNRAS.430.2680M}. The redshift of the DLAs are 
	restricted to $2 < z < 4$, with a median redshift of $z = 2.5$; this redshift is assumed in the model. The DLA parameters are the fiducial parameters of the model in previous sections; Equation \eqref{eq:fidmodel}. The parameters of the $[M/H]-M_v$ distribution are chosen by a maximizing 
	the likelihood: $(\alpha_M, \alpha_v,\beta_M, \sigma_{[M/H]})  = (0.47,0.08,-1.34,0.3)$.
	\emph{Right}: observed and predicted metallicity distribution. The red 
	line shows the same model as the red contours on the left. The black dotted and green dashed lines illustrate variations in the slope of the $[M/H]-M_v$ relation: $(\alpha_M, \alpha_v, \beta_M) = (0.37,0.08,-1.34), ~ (0.57,0.08,-1.33)$.}
	\label{fig:dNdXZ}
\end{figure*}

Figure \ref{fig:dNdXZ} (left) shows the 68-95-99\% contours in the $[M/H] - v_w$ plane, 
along with the data of \citet{2013MNRAS.430.2680M}. The redshift of the DLAs are 
restricted to $2 < z < 4$, with a median redshift of $z = 2.5$; this redshift is assumed in the model. The DLA parameters are the fiducial parameters of the model in previous sections; Equation \eqref{eq:fidmodel}. The parameters of the $[M/H]-M_v$ distribution are chosen by maximizing 
the likelihood: $(\alpha_M, \alpha_v,\beta_M, \sigma_{[M/H]})
 = (0.47,0.08,-1.34,0.3)$. The observed distribution is slightly narrower but otherwise reasonably 
 described by the theoretical distribution. The mean theoretical $[M/H] - v_w$ relation 
 (dashed red line) passes slightly above the binned mean of the observed distribution (black points
 with 1-$\sigma$ SEM error bars).
 
Figure \ref{fig:dNdXZ} (right) shows the observed and predicted metallicity distribution. The red 
line shows the same model as the red contours on the left. The black dotted and green dashed lines illustrate variations in the slope of the $[M/H]-M_v$ relation: $(\alpha_M, \alpha_v, \beta_M) = (0.37,0.08,-1.34), ~ (0.57,0.08,-1.33)$.

An investigation of the parameter space $(\alpha_M, \alpha_v,\beta_M, \sigma_{[M/H]})$ using the Metropolis-Hastings algorithm shows that $\alpha_M$ and $\alpha_v$ are degenerate. A second, smaller peak in the probability distribution occurs at $(\alpha_M, \alpha_v) = (0.27,1.5)$. With no observational constraints on the halo mass of individual DLAs, such degeneracy is not surprising. The major peak at $(\alpha_M, \alpha_v) = (0.47,0.8)$ has a half-width $\Delta \alpha_M = 0.1$, $\Delta \alpha_v = 0.4$. In particular, the case $\alpha_v = 0$ is not strongly ruled out. In summary, Figure \ref{fig:dNdXZ} shows that DLA metallicities can be accounted for (if not predicted) in our model with a mean DLA halo mass-metallicity relationship with a slope of $0.47 \pm 0.1$. At a particular halo mass, metallicity changes little with velocity width.

\begin{figure}
 \centering
	\includegraphics[width=0.45\textwidth]{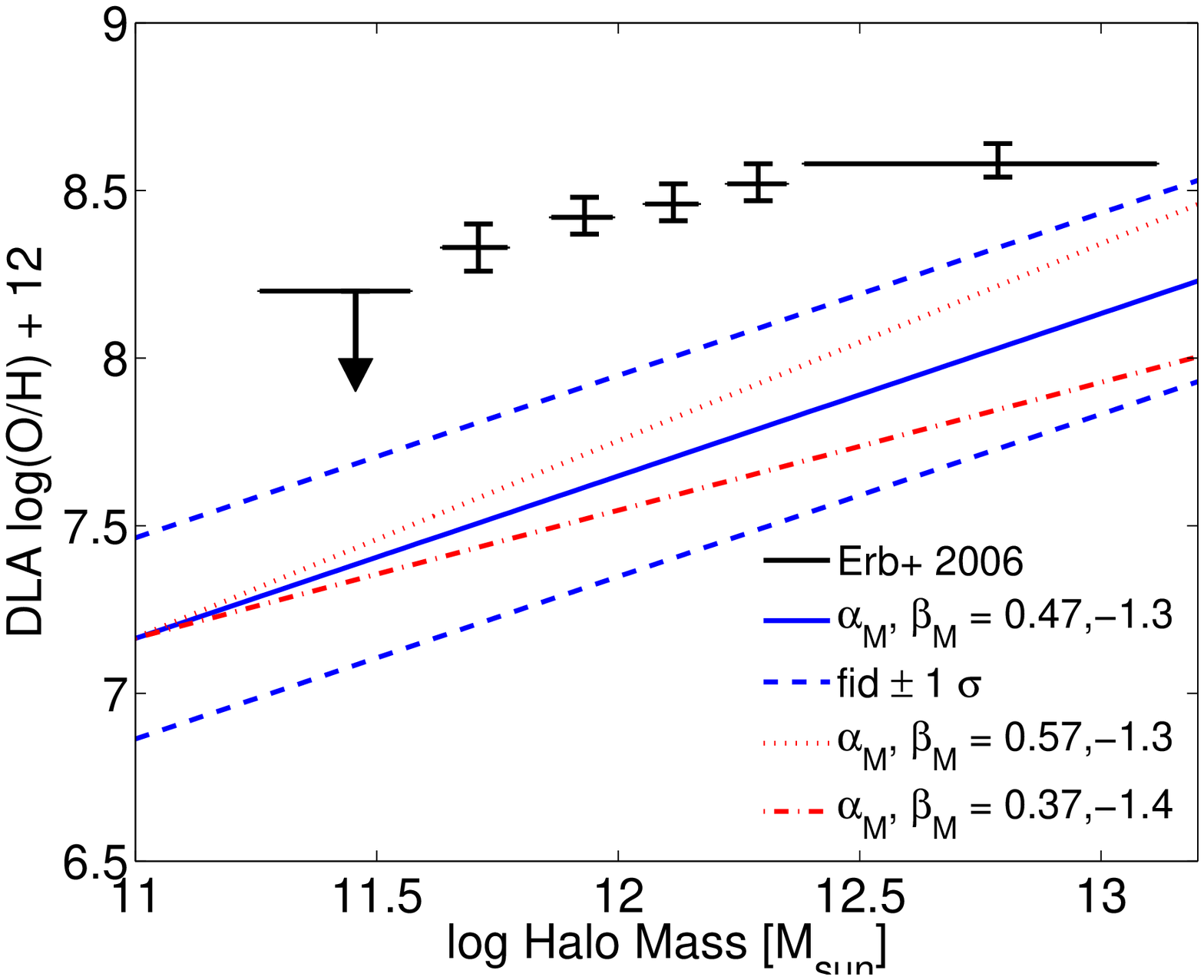}
	\caption{Galaxy and DLA mass-$O/H$ relation. The black points with error bars is the galaxy stellar mass-$O/H$ relation of \citet{2006ApJ...644..813E}, with the conversion from stellar to halo mass from \citet{2013ApJ...770...57B}, as noted above. The solid blue line shows the fiducial DLA mass-metallicity relationship of this section, $(\alpha_M, \beta_M) = (0.47,-1.34)$, with the dashed blue lines showing the $\pm 1 \sigma_{[M/H]}$ Gaussian spread around the mean relation (Equation \ref{eq:Zmass}). The red dotted and dot-dashed lines show the 1-sigma variations in the mean slope $\alpha_M,$, taking into account the degeneracy between $\alpha_M$ and $\beta_M$. The DLAs are significantly more metal-poor at all masses than luminosity-selected galaxies, and even more so at small masses.} \label{fig:Zmass}
\end{figure}

At $z \sim 2.2$, \citet{2006ApJ...644..813E} found that oxygen abundance $[O/H]$ increases with stellar mass. To relate this result to our mass-metallicity relationship, we need to connect DLA metallicity to [O/H], and galaxy stellar mass to halo mass. For the latter relation, we turn to the work of \citet{2013ApJ...770...57B}, who match observed galaxies to haloes to constrain the galaxy-halo relation. The derived halo mass-$[O/H]$ relation is plotted in Figure \ref{fig:Zmass}.

To connect DLA metallicity to $O/H$, we first connect oxygen to iron. \citet{2008MNRAS.385.2011P} published $O/H$ and $Fe/O$ ratios for 33 DLAs with a median redshift of 2.5. The $O/H-Fe/H$ relation can be fit by the following function,
\begin{equation}
[Fe/H] = 0.97 ~ (\log(O/H) + 12) - 8.8 ~,
\end{equation}
where square brackets indicate a logarithmic abundance normalised to solar values; such values are taken from \citet{2009ARA&A..47..481A}. The data vary by $\sigma \approx 0.17$ around this relation. Finally, \citet{2012ApJ...755...89R} note that metallicities derived from Fe include an $\alpha$-enhancement correction, $[M/H] = [Fe/H]+0.3$, giving a final $O/H$-metallicity relation of,
\begin{equation}
[M/H] = 0.97 ~ (\log(O/H) + 12) - 8.5
\end{equation}
Combining this relation with Equation \eqref{eq:Zmass} gives the $O/H$-halo mass relation for our model. It is shown in Figure \ref{fig:Zmass}.

The black points with error bars is the galaxy stellar mass-$O/H$ relation of \citet{2006ApJ...644..813E}, with the conversion from stellar to halo mass from \citet{2013ApJ...770...57B}, as noted above. The solid blue line shows the fiducial DLA mass-metallicity relationship of this section, $(\alpha_M, \beta_M) = (0.47,-1.34)$, with the dashed blue lines showing the $\pm 1 \sigma_{[M/H]}$ Gaussian spread around the mean relation (Equation \ref{eq:Zmass}). The red dotted and dot-dashed lines show the 1-sigma variations in the mean relation, taking into account the degeneracy between $\alpha_M$ and $\beta_M$.

The metallicities measured in absorption in DLAs are significantly lower at all masses than the metallicities measured from the emission of luminosity-selected galaxies. At the same halo mass, the typical difference $\Delta \log(O/H) \sim 1$ is as expected for the metallicity difference between DLAs and LBGs \citep[][Figure 11]{2006fdg..conf..319P}, consistent with two differences between the populations. First, luminosity-selected galaxies are expected to be a brighter, more evolved, and higher star-forming population. Secondly, DLA lines-of-sight are cross-section selected and so will preferentially probe the outer regions of the galaxy \citep[$\sim 4$ kpc][]{2008MNRAS.390.1349P}, while LBG metallicities will tend to probe the central, star-forming region. This is consistent with the strong metallicity gradient ($-0.27 \pm 0.05$  kpc$^{-1}$) observed (via gravitational lensing) in a $z = 2$ galaxy \citep{2010ApJ...725L.176J}. Note, however, that the nine galaxies (mostly at $z < 1.5$ of \citet{2012MNRAS.426..935S} show a shallower average metallicity gradient, in which case the large difference between the metallicities of emission- and absorption-selected galaxies may be another consequence of the velocity width-bias tension, highlighted previously.

\section{Discussion and Conclusions} \label{s:Discussion}

We have investigated whether our previous modelling of the
physical properties of DLAs and faint \lya emitters can account for
the bias parameter estimated by BOSS for DLAs at $z\sim 2.3$. 
We have found that a model in which the fraction of neutral hydrogen in DM haloes drops 
sharply below $\sim 35$ \kmsec reproduces the observed value of the
DLA bias, in broad agreement with the properties of DLAs we derived from their
\lya absorption and emission.

As in \citet{2010MNRAS.403..870B}, our model 
puts DLAs in more massive host haloes than, for example, the numerical simulations of
\citet{2008MNRAS.390.1349P}. This implies that stellar feedback in shallow potential
wells is quite efficient at $z \ga 2.3$.

\citet{2012JCAP...11..059F} parameterised the absorption cross-section
of DLAs as a power law $\sigma_\ro{DLA} \propto M_v^\alpha$. They report
that, \emph{if} they fix their minimum halo mass of a DLA at $10^9
\Msol$, their observed value for the DLA bias is best fit by $\alpha
= 1.1 \pm 0.1$. At $z = 2.3$, $10^9 \Msol$ corresponds to $v_{\ro{v}}
= 21 \kmsec$. Such a population of DLAs in small haloes may 
conflict with the observed distribution of DLA velocity width, as
shown in Figure \ref{fig:vw}. \citet{2013arXiv1308.2598B}
investigated the properties of DLAs in semi-analytics models of galaxy
formation. They report that their favoured ``BRj25'' model produces a slope of
$\alpha = 0.91$, which is closest to the slope reported by
\citet{2012JCAP...11..059F}. However, their simulations were limited
to haloes with masses $\ga 10^{9.7}$\Msol, and the predicted bias is
very sensitive to the low-mass cutoff of the DLA cross-section. If we
extrapolate their $\alpha = 0.91$ model to $10^9 \Msol$, the
predicted bias is $\ga 2 \sigma$ below the observed value. Fixing the
slope, the best fit low-mass cutoff is at $10^{10.3} \Msol$, which
corresponds to a \emph{step-function} cutoff at $v_{\ro{v}} = 56 \kmsec$. 
The modelling of \citet{2012JCAP...11..059F} appears therefore to be consistent 
with our conclusion that some physical process needs to effectively
remove or ionize gas in low mass haloes/shallow potential wells.


We use our constrained model to investigate the DLA halo mass-metallicity relation, 
finding $[M/H]_\ro{mean} = (0.47 \pm 0.1) \log \left( M_v / 10^{11} \Msol \right) 
-1.34$, with no significant metallicity-velocity width relation at 
fixed halo mass. Comparison with the galaxy stellar mass-metallicity relation finds that
DLAs are typically $\Delta \log(O/H) \sim 1$ more metal-poor than luminosity-selected 
galaxies at all masses. We interpret this effect as evidence that DLA sightlines probe the 
outer regions of less-evolved galaxies.


Our modelling of DLA properties, updated to account for the
large BOSS DLA bias parameter, suggests that stellar
feedback in shallow potential wells is more efficient than 
realized in many current numerical galaxy formation
models. Efficient feedback in such rather massive haloes appears also
to be suggested by halo abundance matching analyses
 \citep{2013MNRAS.428.3121M,2013ApJ...762L..31B}. 
As many implementations of galactic winds in numerical simulations already 
struggle to be energetically viable, this adds to the growing
consensus that either the physical mechanism behind driving galactic
winds has not yet been correctly realized in numerical simulations of
galaxy formation, or that other physical processes than efficient
outflows are responsible for the rapidly decreasing stellar and \hi
mass fraction in shallow potential wells. Further consolidation and
extension of the redshift range of measurements of the bias of DLA
host galaxies in combination with improved measurements of the
velocity width distribution of the associated metal distribution
based on larger samples should thus provide important bench marks for the
modelling of stellar feedback in galaxy formation.


\section*{Acknowledgements} 
LAB is funded by the SuperScience Fellowships scheme of the Australian Research Council. MGH acknowledges support from the FP7 ERC Advanced Grant Emergence-320596. 
This work was further supported in part by the National Science Foundation 
under Grant no. PHYS-1066293 and the hospitality of the Aspen Center
for Physics. We would like to thank the anonymous referee for their useful suggestions.


\appendix 

\section{The Observed DLA Velocity Width Distribution}

\begin{figure*} \centering
	\begin{minipage}{0.45\textwidth}
		\includegraphics[width=\textwidth]{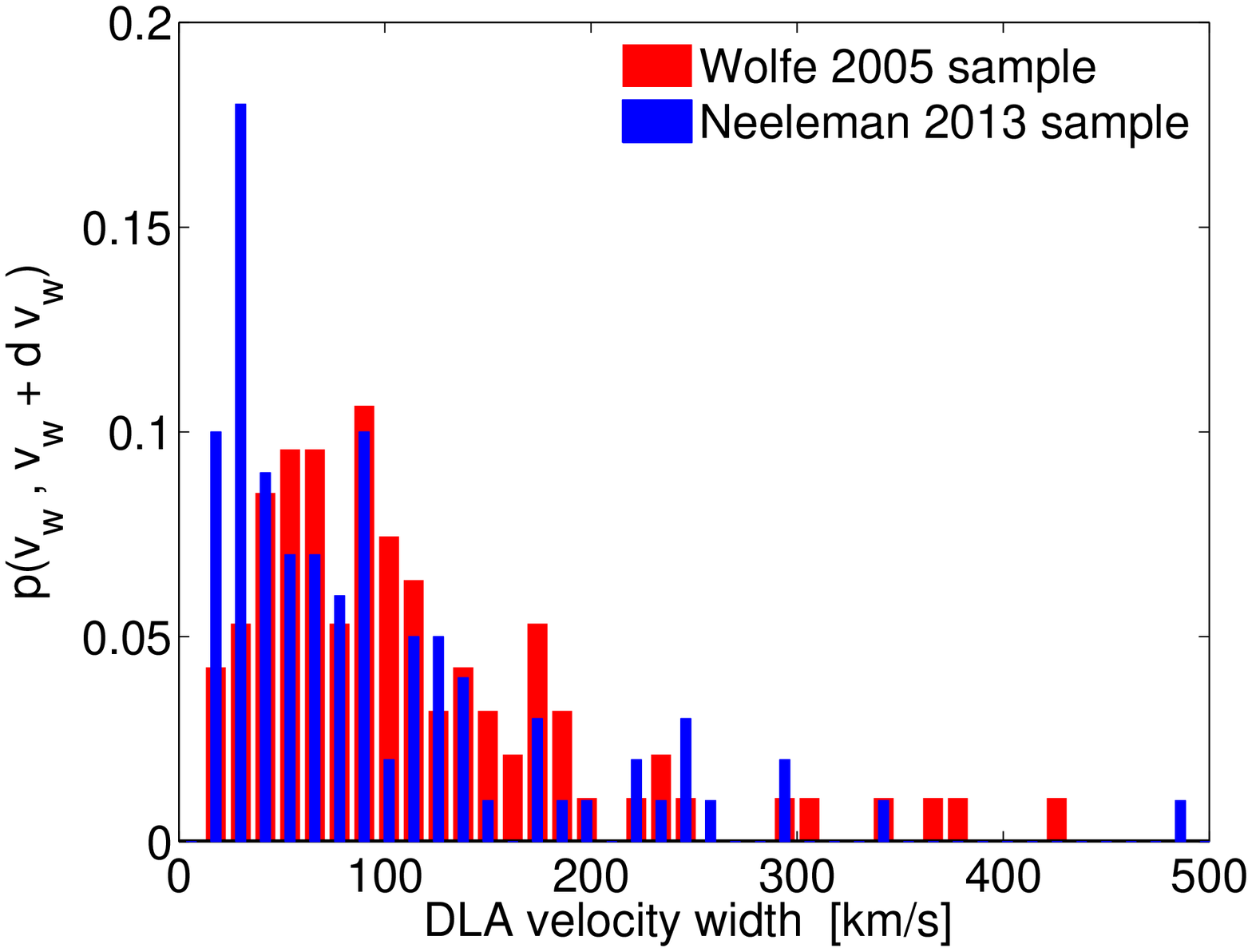}
	\end{minipage}
	\begin{minipage}{0.45\textwidth}
		\includegraphics[width=\textwidth]{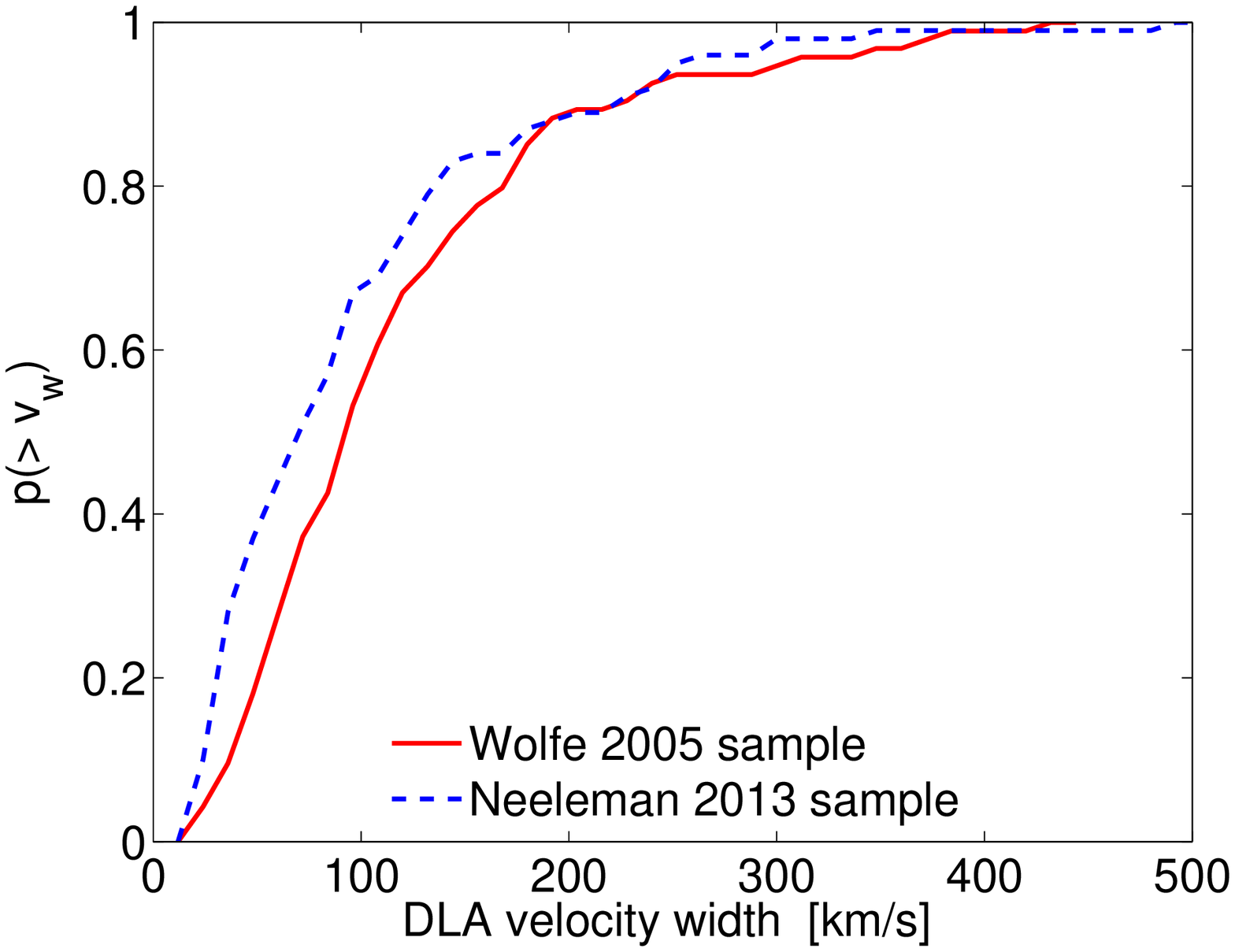}
	\end{minipage}
	\caption{A comparison of the samples of the velocity width of
 the associated low ionization of DLAs from 
 \citet{2005ARA&A..43..861W} and
 \citet{2013ApJ...769...54N}. The left plot shows the
 probability of a DLA having a velocity width $v_w$ in a
 given bin. The right plot shows the cumulative probability
 distribution. There is a considerable difference between the two
 samples: for example, a Kolmogorov-Smirnov test concludes that there is a
 probability of $\sim 1$\% that the two samples are drawn from the
 same underlying population. } \label{fig:vwWN}
\end{figure*}

The velocity width $v_w$ of low-ion metal lines associated with DLAs was
first investigated in detail by \citet{1997ApJ...487...73P} as a probe of 
the kinematic state of the absorbing neutral gas
inside DLAs. The distribution
of $v_w$ has proven to be a challenge for numerical galaxy formation
simulations to reproduce, as noted in the introduction. These
simulations have typically attempted to model the compilation of 94
$v_w$ measurements of \citet{2005ARA&A..43..861W}. Recently,
\citet{2013ApJ...769...54N} released a partially overlapping sample of
100 $v_w$ measurements, all observed with the High Resolution Echelle
Spectrometer (HIRES) on the Keck I 10m telescope. A comparison of the
two samples is shown in Figure \ref{fig:vwWN}.

The left plot shows the probability of a DLA having a velocity width
$v_w$ in a given bin. The right plot shows the cumulative probability
distribution. We note that there is a considerable difference between the two
samples: for example, a Kolmogorov-Smirnov test concludes that there is a
probability of $\sim 1$\% that the two samples are drawn from the
same underlying population.

The samples differ most at the low $v_w$ end, with the \citet{2013ApJ...769...54N}
data set having many more systems with $v_w \lesssim 35 \kmsec$. These are the DLAs that
shed most light on the smallest haloes that are deep enough to hold their
baryons against ejective feedback from supernovae and shield them
from photoionisation. Given that such feedback processes are a major
unknown in galaxy formation simulations, improved observations of the
velocity width distribution of DLAs should thus provide a much needed
stringent test of baryonic physics in low mass haloes/shallow
potential wells.

\bsp 
\label{lastpage}
\end{document}